\documentclass[twocolumn,aps,pra,amsfonts,amssymb]{revtex4-1}
\usepackage{graphicx}
\usepackage{amsmath}
\usepackage{bm}

\newcounter{gg}

\newcommand{\zhat}{\bf \hat{z}}

\newcommand{\pbar}{$\overline{\rm{p}}$~}

\begin{document}

\title{Self-Excitation and Feedback Cooling of an Isolated Proton}

\author{N.\ Guise}
\affiliation{Department of Physics, Harvard University, Cambridge, MA 02138}

\author{J.\ DiSciacca}
\affiliation{Department of Physics, Harvard University, Cambridge, MA 02138}

\author{G.\ Gabrielse}
\email[Email: ]{gabrielse@physics.harvard.edu} \affiliation{Department of Physics, Harvard University, Cambridge, MA
02138}

\date{Submitted to PRL: 7 Dec.\ 2009}

\begin{abstract}     %600 characters including spaces (less than 7 manuscript lines)
The first one-proton self-excited oscillator (SEO) and one-proton feedback cooling are demonstrated.  In a Penning trap with a large magnetic gradient, the SEO frequency is resolved to the high precision needed to detect a one-proton spin flip.  This is after undamped magnetron motion is sideband-cooled to a 14 mK theoretical limit, and despite random frequency shifts (typically larger than those from a spin flip) that take place every time sideband cooling is applied. The observations open a possible path towards a million-fold improved comparison of the \pbar and p magnetic moments.
\end{abstract}


\maketitle

\newcommand{\w}{3.00in}

\newcommand{\SXOSchematicSimpleFigure}{
\begin{figure}[htbp!]
\includegraphics*[width=0.85\columnwidth]{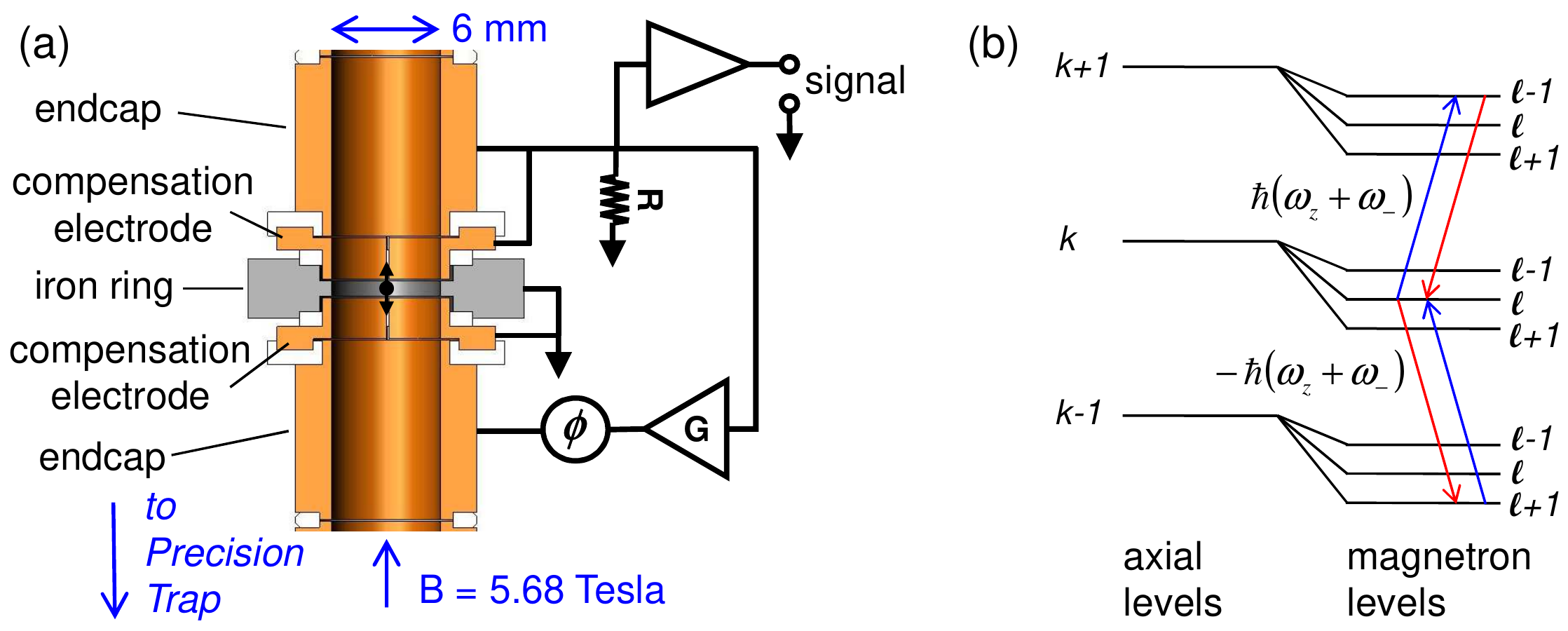}
\caption{(a) Penning trap electrodes and radiofrequency schematic for feedback cooling and self-excitation of the proton
axial motion.  The feedback has gain $G$ and a phase shifted by $\phi$. (b) Energy levels and transitions (arrows) involved
in axial sideband cooling of proton magnetron motion.} \label{fig:SXOSchematicSimple}
\end{figure}
}

\newcommand{\PeakDipResolutionFigure}{
\begin{figure}[htbp!]
\includegraphics*[width=1.0\columnwidth]{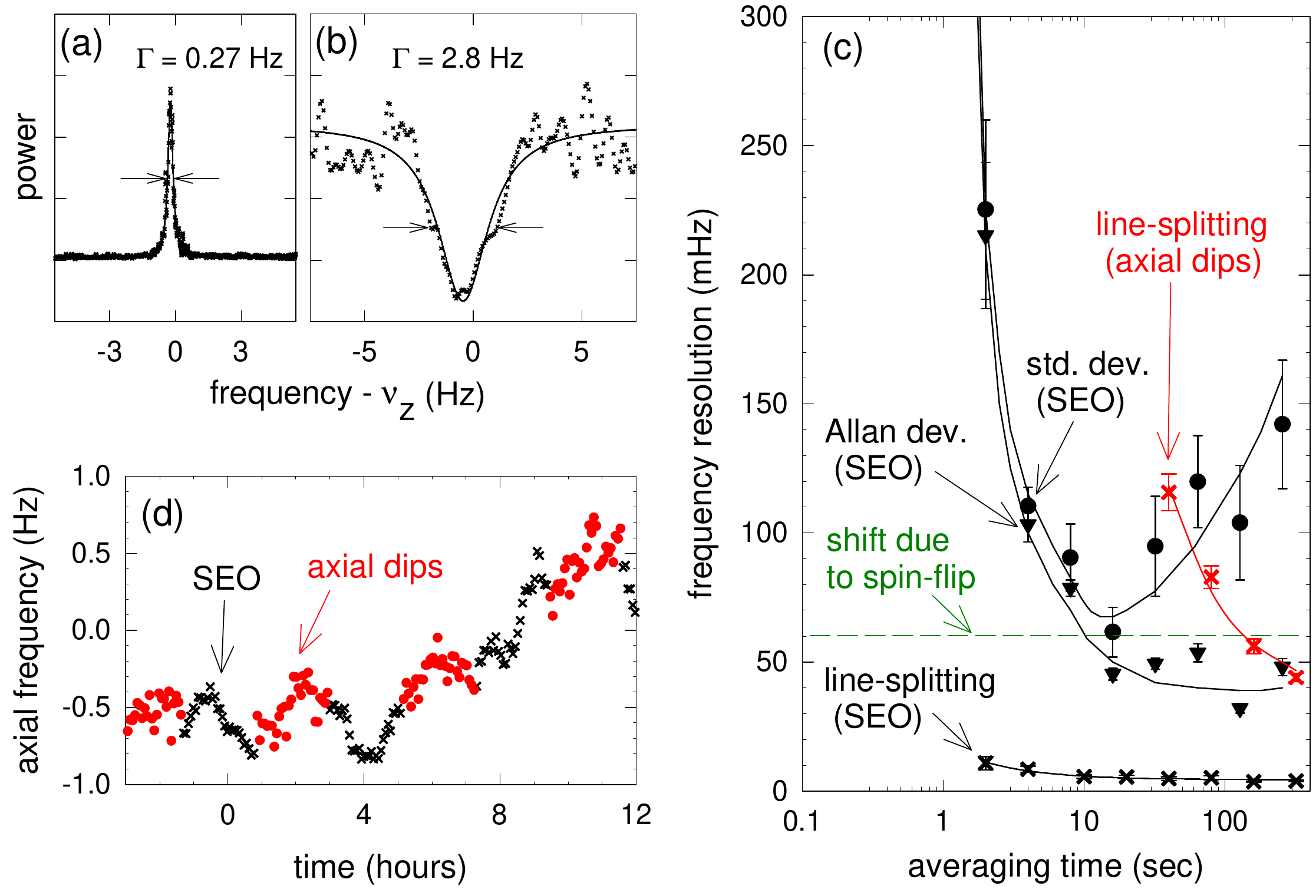}
\caption{SEO peak (a) and noise dip (b) for 160 s of averaging.  (c) Frequency resolution achieved with a single average of
an SEO peak (black x) and noise dip (red x), with the standard deviation (black points) and Allan deviation (black
triangles) of averaged SEO measurements.  (d) Drift of 256 s averages over sixteen night time hours.}
\label{fig:PeakDipResolution}
\end{figure}
}

\newcommand{\SEOFigure}{
\begin{figure}[htbp!]
\includegraphics*[width=1.0\columnwidth]{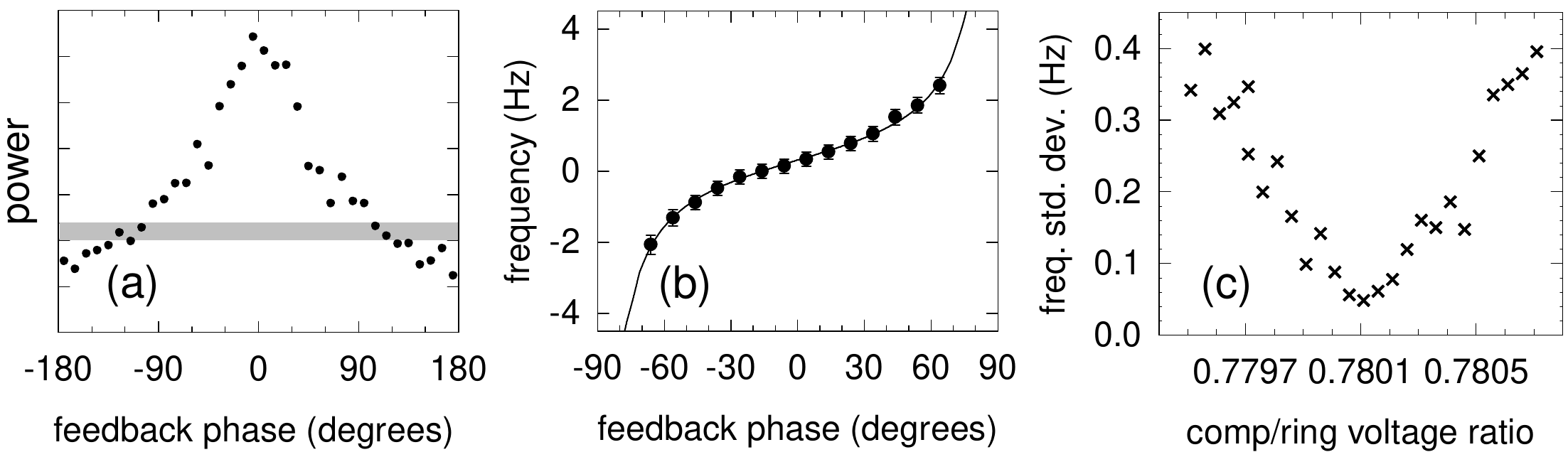}
\caption{(a) SEO signal strength vs.\ feedback phase.  (b) Measured axial frequency vs.\ feedback phase  (points) fit to the
expected Eq.~\ref{eq:SXOPhaseDependence}. (c) Tuning for optimal SEO stability by adjusting trap anharmonicity.}
\label{fig:SEO}
\end{figure}
}

\newcommand{\MagnetronHistogramFigure}{
\begin{figure}[htbp!]
\includegraphics*[width=1.0\columnwidth]{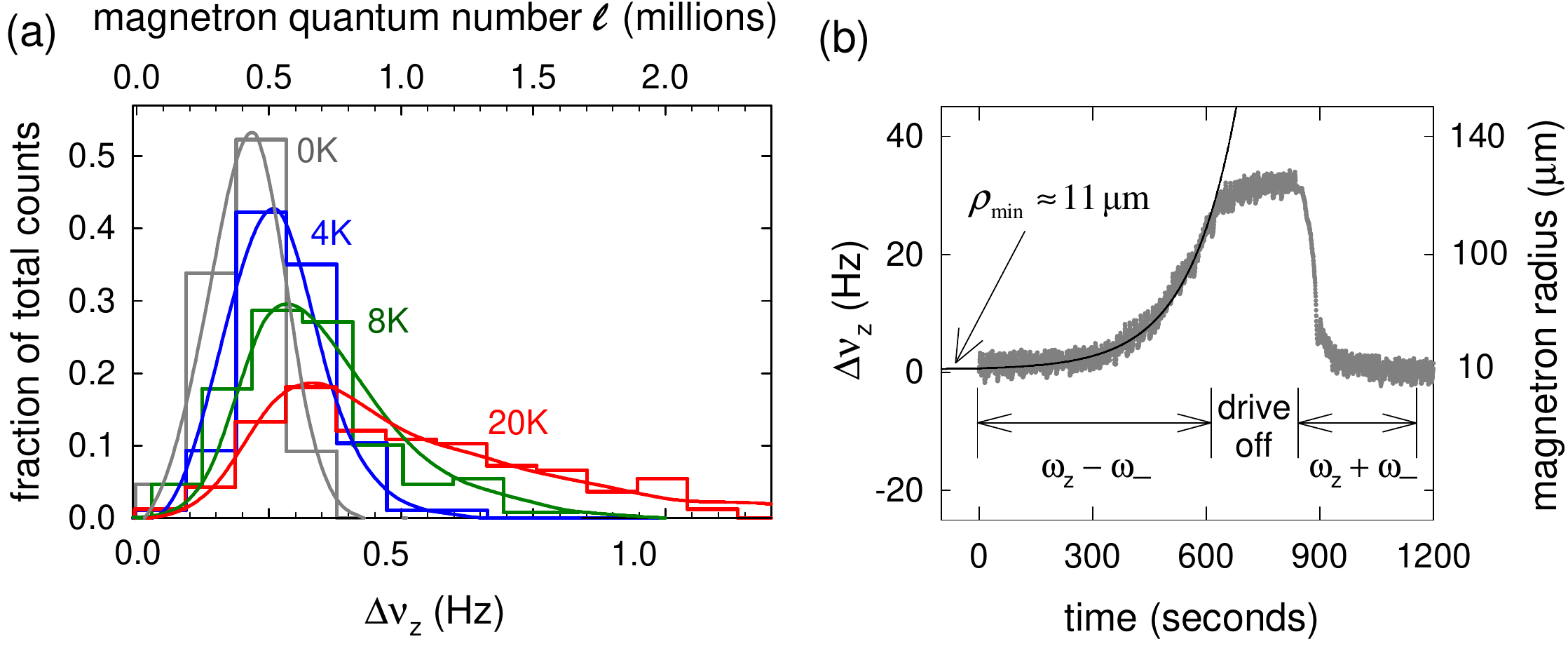}
\caption{(a) Histograms of magnetron states after no sideband cooling (gray), and produced by sideband cooling using
feedback cooling (blue), no feedback (green), and with feedback heating (red).  Solid curves are convolutions of the gray
Gaussian resolution function and Boltzmann distributions at the specified $T_z$. (b) The magnetron radius increase from a
sideband drive at $\omega_z-\omega_-$ is fit to an exponential and extrapolated back to an initial magnetron radius.}
\label{fig:MagnetronHistogram}
\end{figure}
}

\newcommand{\FeedbackCoolingFigure}{
\begin{figure}[htbp!]
\includegraphics*[width=1.0\columnwidth]{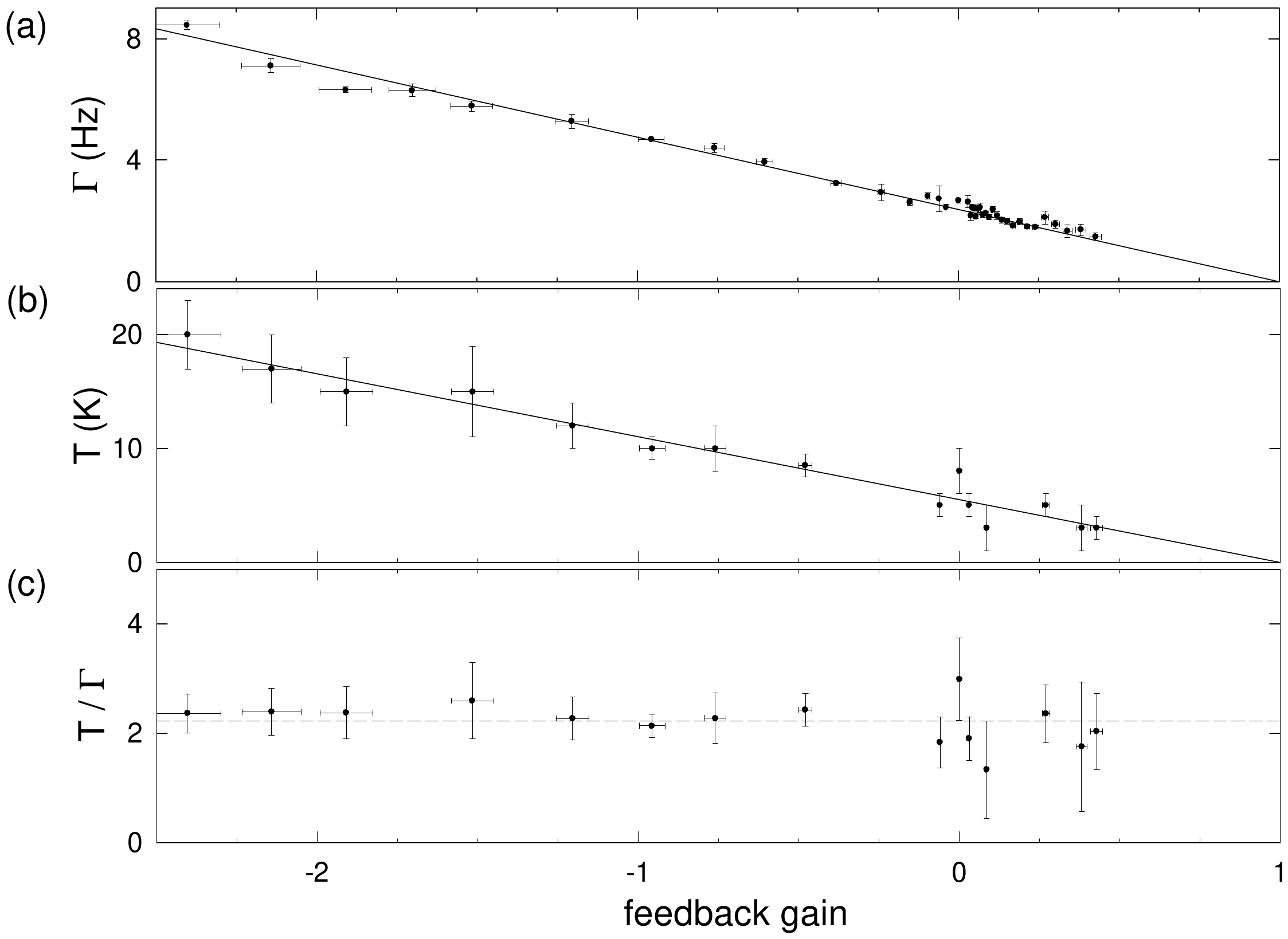}
\caption{Measured damping widths (a), temperatures (b), and their ratios (c) as a function of the feedback gain, $G$.}
\label{fig:FeedbackCooling}
\end{figure}
}

The demonstration of a one-electron self-excited oscillator (SEO) \cite{SelfExcitedOscillator} and one-electron feedback cooling \cite{FeedbackCoolingPRL} eventually led to greatly improved measurements of the electron magnetic moment and the fine structure constant \cite{HarvardMagneticMoment2008}.  An electron spin flip caused the SEO oscillation frequency to shift observably because of a magnetic gradient added to the Penning trap that held the electron.

This Letter reports the first one-proton SEO and the first one-proton feedback cooling.  The SEO frequency is resolved at the very precise level needed to observe a spin flip of a single \pbar or p -- opening a possible way to compare the \pbar and p magnetic moments \cite{SelfExcitedOscillator,QuintAntiprotonAspirations} a million times more precisely than current comparisons.  To compensate for the \pbar and p moments being 650 times smaller than that of an electron, these observations took place in a smaller trap with a much larger gradient than sufficed for observing spin flips of free \cite{HarvardMagneticMoment2008} and bound \cite{MainzSummary2006} electrons.  Feedback cooling promises to narrow the spin resonance linewidth and increase the transition rate.

Effective sideband cooling \cite{DehmeltMagneticBottle,Review} is essential for these measurements, to center the p or \pbar so that electric and magnetic
anharmonicities do not mask the tiny frequency shifts that would signal
spin flips.  Sideband cooling is widely utilized not only in precision
measurements of trapped particles, but also for the manipulation of
qubits \cite{Entanglement2008}, the most precise clocks \cite{CompareAlAndHgIonClocks2008}, the cooling of neutral atoms
in lattices \cite{SidebandCoolingOfNeutralAtoms1998}, and the cooling of mechanical resonators \cite{SidebandCoolingMechanicalResonator}.  The one-proton SEO and feedback cooling make it possible to directly
investigate the outcome of sideband cooling to the low-temperature,
theoretical limit \cite{WinelandFrequencyDivision,Review}, for a particle that has no internal degrees
of freedom available for cooling.  The observed distribution of proton
magnetron states is at the 14 mK theoretical limit.  Previous such
investigations with an electron \cite{VanDyckMagnetronCoolingLimit,Review} reached only an energy 400
times higher than the desired low temperature limit, and
investigations with an ion probed temperatures orders of magnitude
higher \cite{QuintDoublePenningTrap,OneIonTemperature}.   A complication due to the strong magnetic gradient required is that an application of sideband cooling
randomly shifts the SEO frequency by an amount larger than the shift
from a spin flip on average.

A single proton is suspended in a vertical $B = 5.68$ Tesla field at the center of a cylindrically symmetric trap (Fig.~\ref{fig:SXOSchematicSimple}) -- stacked rings with a 3 mm inner radius.  The electrodes and surrounding vacuum container are cooled to 4.2 K by a thermal connection to liquid helium.  Cryopumping of the closed system made the vacuum better than $5 \times 10^{-17}$ Torr in a similar system \cite{PbarMass}, so collisions are not important. Appropriate potentials applied to copper electrodes (with an evaporated gold layer) in an open-access geometry \cite{OpenTrap} make a very good electrostatic quadrupole near the trap center, while also maintaining an open access to the trap interior from either end.

\SXOSchematicSimpleFigure

The proton's circular cyclotron motion is perpendicular to {\bf B}, with a frequency $\omega_+ /(2 \pi) = 79.5$ MHz that is
slightly modified by the electrostatic potential.  The proton also oscillates parallel to {\bf B} at $\omega_z/(2 \pi) =
\,553$ kHz.  Because the potential is not a perfect quadrupole, this frequency depends slightly upon oscillation amplitude,
$A$, with $\omega_z(A) \approx \omega_z$.  The proton's third motion is a circular magnetron motion, also perpendicular to
{\bf B}, at the much lower frequency, $\omega_- /(2 \pi) = 1.9$ kHz.

To couple the proton spin moment and the magnetron state that is the outcome of sideband cooling to the measurable $\omega_z(A)$, the trap's central
ring electrode is made of saturated iron (unlike the copper endcap and compensation electrodes above and below). The
extremely large magnetic bottle gradient,
\begin{equation}
\Delta {\bf B} = \beta_2 [(z^2-\rho^2/2){\bf \hat{z}} - z\rho \boldsymbol{\hat{\rho}}]
\end{equation}
with $\beta_2 = 7.8 \times 10^4$ T/m$^2$, is 51 and 8 times larger than what was used to measure free \cite{HarvardMagneticMoment2008} and bound \cite{QuintDoublePenningTrap,MainzSummary2006} electron magnetic moments. The bottle reduces the field within the trap by $0.47$ T (8 \%).

The axial frequency $\omega_z(A)$ depends primarily on the strength of the $z^2$ term in the electrostatic quadrupole. A
magnetic moment $\mu \zhat$ (from circular cyclotron or magnetron motions, or from spin) adds a term going as $\mu z^2$ to
the trapping potential, shifting $\omega_z(A)$ by
\begin{equation} \frac{\Delta \omega_z}{\omega_z} \approx \frac{\hbar
 \beta_2}{2 m_p \omega_- |B|} \left( n + \frac{1}{2} + \frac{g_p m_s}{2} + \frac{\omega_-}{\omega_+} (\ell + \frac{1}{2})
\right). \label{eq:FrequencyShift}
\end{equation}
The magnetic moments from cyclotron and magnetron motion go as $n$ and $\ell$.   The 553 kHz axial frequency shifts by $21$
mHz per cyclotron quantum ($200$ Hz per $\mu$m for our typical cyclotron radius), and by $0.5$ $\mu$Hz per magnetron quantum
(40 mHz per $\mu m$ at our typical magnetron radius).  A proton spin flip will cause a 60 mHz shift.

The frequency of the current induced to flow through $R$ in Fig.~\ref{fig:SXOSchematicSimple} by the  axial oscillation is measured to determine $\omega_z(A)$. The voltage across $R = 25~\rm{M}\Omega$ is Fourier transformed after an amplifier that uses a high electron mobility transistor (HEMT) with  good thermal connection to 4.2 K. The $I^2R$ loss for the induced current
going through $R$ gives an axial damping time $\gamma_z^{-1} = 60 $ ms.

$R$ represents losses in an LC tuned circuit resonant at $\omega_z$. The losses are minimized to maximize $R$, the observed
signal and the damping rate. Varactors tune the circuit and its matching to the HEMT. A superconducting inductor with $L =
2.5$ mH cancels the reactance of the trap capacitance, leaving $R = Q \omega_z L$.  The circuit's quality factor is tuned to
$Q = 3000$ with the trap electrodes attached. (With the amplifier connected only to the endcap, $Q = 5000$. With a capacitor
substituted for the trap electrodes, $Q\ge 20,000$.)  The proton's cyclotron and magnetron motions in this trap are not
damped.

A nearly identical ``precision trap'' is just below the trap in Fig.~\ref{fig:SXOSchematicSimple}. A detection circuit
resonant at $\omega_+/(2\pi) = 86.5$ MHz, attached across halves of a copper ring electrode, damps this motion in
$\gamma_+^{-1}=10$ min.  (This could be three times faster with better amplifier tuning.) An axial amplifier detects and
damps the axial motion.

A single proton is isolated in the second trap using a relativistic method we developed earlier with antiprotons
\cite{PbarMassPRL1ppb}. An H atom is ionized in the trap by an $e^-$ beam from a sharp field emission point. Strong driving
forces applied at the axial frequencies of unwanted ions keep them from loading. A strong pulse of cyclotron drive produces
one-proton cyclotron resonances that differ in frequency because of differing cyclotron energies and relativistic mass
shifts of $\omega_+$. The trap potential is temporarily reduced until the signal from only one proton remains. The cyclotron
energy of the remaining proton damps until its radius is less than the $0.5$ $\mu m$ average for a 4.2 K distribution. After
magnetron cooling, the proton is transferred into the trap of Fig.~\ref{fig:SXOSchematicSimple} by adjusting applied
potentials to make an axial potential well that moves adiabatically from the lower to the upper trap.

The proton axial oscillation whose frequency is to be measured satisfies the equation of motion,
\begin{equation}
  \ddot{z} + \gamma_z\dot{z} + [\omega_z(A) + \Delta \omega_z]^2 z
  = F_d(t)/m.
  \label{eq:EquationOfMotion}
\end{equation}
A driving force $F_d(t)$ is added to the restoring force (from the electrostatic quadrupole and the magnetic bottle), and to
the damping force $-m\gamma_z \dot{z}$ (from the loss in $R$).

With no feedback, $F_d$ is the Johnson noise from the resistor that is then amplified and detected.  The proton's axial
oscillation shorts this noise \cite{ElectronCalorimeter}, making a dip in the noise power spectrum
(Fig.~\ref{fig:PeakDipResolution}b) whose half width is $\gamma_z$.

\PeakDipResolutionFigure

The axial frequency is determined to higher precision using the better signal-to-noise and narrower signal width of a
one-proton SEO (Fig.~\ref{fig:PeakDipResolution}a). The one-particle SEO, realized previously only with an electron
\cite{SelfExcitedOscillator}, is realized by adjusting the amplitude and phase of the amplified induced signal and feeding
this back to the other side of the trap as a driving force on the proton. The feedback produces a force $F_d(t) \sim m G
\gamma_z \dot z$ with feedback gain $G$. Self-excitation occurs, in principle, when the feedback cancels the damping at
$G=1$. Noise causes amplitude diffusion and energy growth, however, and $G$ slightly different from unity will either
decrease or increase $A$ exponentially. A stable and useful SEO thus requires limiting the amplitude to some value $A_o$.
Here a digital signal processor (DSP) chip Fourier transforms the signal to determine $A$, and makes $G=1 + a(A-A_o)$
\cite{SelfExcitedOscillator}.

An axial oscillation $z(t)=A \cos(\omega t)$ generates a feedback force $F_d(t)=-\omega A G m \gamma_z \sin(\omega t+\phi)$,
when a phase shift $\phi$ is introduced (Fig.~\ref{fig:SXOSchematicSimple}).  Inserting in Eq.~\ref{eq:EquationOfMotion}
yields
\begin{eqnarray}
\label{eq:SXOPhaseEquations}
G \cos(\phi)&=&1 \\
G\, \omega \, \gamma_z \sin(\phi)&=& \omega^2 - \left[\omega_z(A)\right]^2
\end{eqnarray}
For $(\gamma_z/\omega_z) \tan(\phi) << 1$, the SEO thus depends on $\phi$ as
\begin{equation}
\label{eq:SXOPhaseDependence} \omega(A,\phi) \approx \omega_z(A) +
\frac{\gamma_z}{2}tan(\phi) \; .
\end{equation}
With positive feedback, and a feedback phase adjusted to  optimize the signal (Fig.~\ref{fig:SEO}a), the measured SEO
frequency as a function of feedback phase fits well to Eq.~\ref{eq:SXOPhaseDependence} (Fig.~\ref{fig:SEO}b).  The scatter
in repeated frequency measurements (Fig.~\ref{fig:SEO}c) is reduced when the trapping potential is tuned to make the best
possible electrostatic quadrupole.

\SEOFigure

Sideband cooling, a method to radially center the proton, is especially important given that the magnetic field changes significantly as a function of radial position. A sideband cooling drive at $\omega_z+\omega_-$, applied across the halves of a compensation electrode
(Fig.~\ref{fig:SXOSchematicSimple}) to produce an $zy$ potential gradient \cite{Review}, makes the transitions between axial
states $k$ and magnetron states $\ell$ indicated by arrows in Fig.~\ref{fig:SXOSchematicSimple}b. The probability
$P_{k,\ell}$ satisfies
\begin{equation}
0=-P_{k,\ell}\,(\Gamma_+ + \Gamma_-) + P_{k-1,\ell+1}\,\Gamma_- + P_{k+1,\ell-1}\,\Gamma_+. \label{eq:RateEquation}
\end{equation}
This steady-state rate equation has rates that depend on axial and magnetron raising and lowering operators \cite{Review},
\begin{eqnarray}
\Gamma_+ &\sim& |\!<\!k+1,\ell-1|a_z^\dagger \,a_- |k,\ell\!>\!|^2 \sim (k+1)\ell \\
\Gamma_- &\sim& |\!<\!k-1,\ell+1|a_z \, a_-^\dagger |k,\ell\!>\!|^2 \sim k(\ell +1).
\end{eqnarray}
The axial distribution remains a Boltzmann distribution due to its coupling to the detection resistor, a reservoir at
temperature $T_z$, so that $P_{k,\ell} = p_\ell\,\exp{[-k\hbar\omega_z/(k_B T_z)]}$. The solution to
Eq.~\ref{eq:RateEquation} is a magnetron distribution,
\begin{equation}
p_\ell \sim \exp{[-\ell\hbar\omega_-/(k_B T_m)]},\label{eq:MagnetronDistribution}
\end{equation}
where the effective magnetron temperature is
 $T_m=T_z \omega_-/\omega_z$.
The theoretical cooling limit
\cite{WinelandFrequencyDivision,Review},
\begin{equation}
k_B T_m =  \langle -E_{mag}\rangle = \frac{\omega_-}{\omega_z} \langle E_z \rangle \,=  \frac{\omega_-}{\omega_z} \, k_B T_z,
\label{eq:SidebandCoolingLimit}
\end{equation}
comes from evaluating the average of the magnetron energy, $\langle E_{mag} \rangle \,= \sum_\ell p_\ell \, E_\ell$, with
$E_\ell = -(\ell + 1/2)\hbar \omega_-$. Decreasing $\ell$ increases the potential energy on a radial hill, while decreasing
the much smaller kinetic energy, so $E_{\ell}$ is negative. Ref.~\cite{Review} extends the theoretical argument and limit to
off-resonant sideband cooling drives.

The outcome of sideband cooling in the trap of Fig.~\ref{fig:SXOSchematicSimple}a
is investigated with a three-step
sequence. First,  the axial energy is either left in equilibrium with the detection resistor or modified using feedback.
Second, a sideband cooling drive at $\omega_z+\omega_-$ is applied and then turned off. Third, the SEO is started and the
axial frequency measured. Each application of sideband cooling produces a measurably different magnetron state and
$\omega_z(A)$, so the sequence is repeated to make histograms of measured axial frequencies.

The gray histogram and Gaussian fit in Fig.~\ref{fig:MagnetronHistogram}a show the scatter for repeated measurements of
$\omega_z(A)$ taken with no sideband cooling drive (i.e.\ no change in magnetron radius) and no feedback (i.e.\ no change in
$T_z$).

\MagnetronHistogramFigure

Sideband cooling with no feedback broadens the gray into the green histogram (Fig.~\ref{fig:MagnetronHistogram}a). A convolution (green curve) of Eq.~\ref{eq:MagnetronDistribution} with  $T_m=30$ mK (corresponding to $T_z = 8\pm 2$ K) and the gray Gaussian resolution function fits the measured histogram when Eq.~\ref{eq:FrequencyShift} is used to convert magnetron energy to axial frequency shift. The axial temperature is reasonably higher than the $T_z=5.2$ K we realized with one electron in a 1.6 K apparatus \cite{FeedbackCoolingPRL}.

Feedback changes the measured $T_z$ as predicted, from $T_{z0}$ at $G$=0 to $T_z(G) = (1-G)T_{z0}$
(Fig.~\ref{fig:FeedbackCooling}b), increasing our confidence in this new way to measure the axial temperature. The
corresponding damping widths also change just as predicted, from $\Gamma_{z0}$ to $\Gamma_z(G) = (1-G)\Gamma_{z0}$
(Fig.~\ref{fig:FeedbackCooling}a). The ratios in Fig.~\ref{fig:FeedbackCooling}c are constant, consistent with the
fluctuation-dissipation theorem.

\FeedbackCoolingFigure

Feedback cooling of the axial motion to $T_z=4$ K narrows the distribution of magnetron states (blue histogram in Fig.~\ref{fig:MagnetronHistogram}a) such that the effective magnetron temperature is $T_m=14$ mK. Feedback cooling from $T_z = 8$ to $4$ K seems plausibly higher than the cooling from 5.2 to 0.85 K \cite{FeedbackCoolingPRL} achieved with one electron in a 1.6 K apparatus.  Feedback heating of the axial motion to $T_z=20$ K broadens the distribution (red histogram in Fig.~\ref{fig:MagnetronHistogram}a).

A check on the magnetron orbit size produced by sideband cooling comes from expanding the orbit size exponentially with a sideband heating drive at $\omega_z - \omega_-$ \cite{VanDyckMagnetronCoolingLimit,Review}. Each trial (e.g. Fig.~\ref{fig:MagnetronHistogram}b) is extrapolated to determine the radius at the start of the heating.  Averaging the initial radii from 200 trials gives  $11\pm2~\mu\rm{m}$, consistent with $T_z=8\pm2$ K from
Fig.~\ref{fig:MagnetronHistogram}a, and hence with the theoretical cooling limit (Eq.~\ref{eq:SidebandCoolingLimit}). We do
not understand the earlier electron observations \cite{VanDyckMagnetronCoolingLimit,Review} but note that progress has been made in the detection electronics that sets $T_z$.

Fig.~\ref{fig:PeakDipResolution}c shows that for the first time in a strong magnetic gradient, the resolution achieved in measuring $\omega_z$ is comparable to the 60 mHz resolution needed to observe a proton spin flip (for a 16 s averaging time). The resolution is much better for a single SEO measurement (black x) than for a single dip measurement (red x). Repeated SEO frequency measurements have a standard deviation (black points) and an Allan deviation (black triangles) that is larger than the precision for a single measurement.  Fluctuations in the trapping potential, mechanical vibrations, temperature variations, and  fluctuating contact potentials are being investigated as possible sources of the scatter.

Sideband cooling is required to minimize the scatter and achieve the 60 mHz resolution. However, Fig.~\ref{fig:MagnetronHistogram}a demonstrates that an application of sideband cooling randomly selects a new magnetron radius and $\omega_z$  that is shifted by much more than this resolution on average. A solution starts with an initial sideband cooling period, after which $\omega_z$ is measured.  An attempt to make a spin flip can be then made for several minutes, during which time unwanted magnetron heating shifts $\omega_z$ typically by about 0.3 Hz/hr (Fig.~\ref{fig:PeakDipResolution}d).  Measuring $\omega_z$ will thus reveal a shift larger or smaller than 60 mHz depending upon whether the spin has or has not flipped. The process can then be repeated, with cooling selecting a new $\omega_z$.

In conclusion, a one-proton self-excited oscillator and one-proton feedback cooling are realized for the first time. A very strong magnetic gradient is added to the Penning trap in which the proton is suspended to make it possible to observe sideband cooling distributions and to investigate the possibility of observing spin flips.
Sideband cooling of the undamped proton magnetron motion to 14 mK is demonstrated, even though every application of sideband cooling shifts the monitored SEO oscillation frequency more than would a spin flip. As an application, the SEO oscillation frequency is resolved at the high precision needed to observe a spin flip of a single \pbar or p, opening a possible new path towards comparing the \pbar and p magnetic moments at a precision higher than current comparisons by six orders of magnitude or more.

Thanks to the NSF AMO program, the AFOSR, the NDSEG, and the Humboldt Foundation for support.

%\bibliography{d:/jerry/shared/synchronize/ggrefs2010}
%\bibliography{c:/users/jerry/shared/synchronize/ggrefs2009}

\begin{thebibliography}{10}%
\makeatletter
\providecommand \@ifxundefined [1]{%
 \ifx #1\undefined \expandafter \@firstoftwo
 \else \expandafter \@secondoftwo
\fi
}%
\providecommand \@ifnum [1]{%
 \ifnum #1\expandafter \@firstoftwo
 \else \expandafter \@secondoftwo
\fi
}%
\providecommand \enquote [1]{``#1''}%
\providecommand \bibnamefont  [1]{#1}%
\providecommand \bibfnamefont [1]{#1}%
\providecommand \citenamefont [1]{#1}%
\providecommand\href[0]{\@sanitize\@href}%
\providecommand\@href[1]{\endgroup\@@startlink{#1}\endgroup\@@href}%
\providecommand\@@href[1]{#1\@@endlink}%
\providecommand \@sanitize [0]{\begingroup\catcode`\&12\catcode`\#12\relax}%
\@ifxundefined \pdfoutput {\@firstoftwo}{%
 \@ifnum{\z@=\pdfoutput}{\@firstoftwo}{\@secondoftwo}%
}{%
 \providecommand\@@startlink[1]{\leavevmode}%
 \providecommand\@@endlink[0]{}%
}{%
 \providecommand\@@startlink[1]{%
  \leavevmode
  \pdfstartlink
   attr{/Border[0 0 1 ]/H/I/C[0 1 1]}%
   user{/Subtype/Link/A<</Type/Action/S/URI/URI(#1)>>}%
  \relax
 }%
 \providecommand\@@endlink[0]{\pdfendlink}%
}%
\providecommand \url  [0]{\begingroup\@sanitize \@url }%
\providecommand \@url [1]{\endgroup\@href {#1}{\urlprefix}}%
\providecommand \urlprefix [0]{URL }%
\providecommand \Eprint[0]{\href }%
\@ifxundefined \urlstyle {%
  \providecommand \doi [1]{doi:\discretionary{}{}{}#1}%
}{%
  \providecommand \doi [0]{doi:\discretionary{}{}{}\begingroup
  \urlstyle{rm}\Url }%
}%
\providecommand \doibase [0]{http://dx.doi.org/}%
\providecommand \Doi[1]{\href{\doibase#1}}%
\providecommand \bibAnnote [3]{%
  \BibitemShut{#1}%
  \begin{quotation}\noindent
    \textsc{Key:}\ #2\\\textsc{Annotation:}\ #3%
  \end{quotation}%
}%
\providecommand \bibAnnoteFile [2]{%
  \IfFileExists{#2}{\bibAnnote {#1} {#2} {\input{#2}}}{}%
}%
\providecommand \typeout [0]{\immediate \write \m@ne }%
\providecommand \selectlanguage [0]{\@gobble}%
\providecommand \bibinfo [0]{\@secondoftwo}%
\providecommand \bibfield [0]{\@secondoftwo}%
\providecommand \translation [1]{[#1]}%
\providecommand \BibitemOpen[0]{}%
\providecommand \bibitemStop [0]{}%
\providecommand \bibitemNoStop [0]{.\EOS\space}%
\providecommand \EOS [0]{\spacefactor3000\relax}%
\providecommand \BibitemShut [1]{\csname bibitem#1\endcsname}%
%</preamble>
\bibitem{SelfExcitedOscillator}%
  \BibitemOpen
  \bibfield{author}{%
  \bibinfo {author} {\bibfnamefont{B.}~\bibnamefont{D'Urso}}, \bibinfo {author}
  {\bibfnamefont{R.}~\bibnamefont{{Van Handel}}}, \bibinfo {author}
  {\bibfnamefont{B.}~\bibnamefont{Odom}}, \bibinfo {author}
  {\bibfnamefont{D.}~\bibnamefont{Hanneke}},\ and\ \bibinfo {author}
  {\bibfnamefont{G.}~\bibnamefont{Gabrielse}},\ }%
  \bibfield{journal}{%
  \bibinfo {journal} {Phys. Rev. Lett.}\ }%
  \textbf{\bibinfo {volume} {94}},\ \bibinfo {pages} {113002} (\bibinfo {year}
  {2005})%
  \bibAnnoteFile{NoStop}{SelfExcitedOscillator}%
\bibitem{FeedbackCoolingPRL}%
  \BibitemOpen
  \bibfield{author}{%
  \bibinfo {author} {\bibfnamefont{B.}~\bibnamefont{D'Urso}}, \bibinfo {author}
  {\bibfnamefont{B.}~\bibnamefont{Odom}},\ and\ \bibinfo {author}
  {\bibfnamefont{G.}~\bibnamefont{Gabrielse}},\ }%
  \bibfield{journal}{%
  \bibinfo {journal} {Phys. Rev. Lett.}\ }%
  \textbf{\bibinfo {volume} {90}},\ \bibinfo {pages} {043001} (\bibinfo {year}
  {2003})%
  \bibAnnoteFile{NoStop}{FeedbackCoolingPRL}%
\bibitem{HarvardMagneticMoment2008}%
  \BibitemOpen
  \bibfield{author}{%
  \bibinfo {author} {\bibfnamefont{D.}~\bibnamefont{Hanneke}}, \bibinfo
  {author} {\bibfnamefont{S.}~\bibnamefont{Fogwell}},\ and\ \bibinfo {author}
  {\bibfnamefont{G.}~\bibnamefont{Gabrielse}},\ }%
  \bibfield{journal}{%
  \bibinfo {journal} {Phys. Rev. Lett.}\ }%
  \textbf{\bibinfo {volume} {100}},\ \bibinfo {pages} {120801} (\bibinfo {year}
  {2008})%
  \bibAnnoteFile{NoStop}{HarvardMagneticMoment2008}%
\bibitem{QuintAntiprotonAspirations}%
  \BibitemOpen
  \bibfield{author}{%
  \bibinfo {author} {\bibfnamefont{W.}~\bibnamefont{Quint}}, \bibinfo {author}
  {\bibfnamefont{J.}~\bibnamefont{Alonso}}, \bibinfo {author}
  {\bibfnamefont{S.}~\bibnamefont{Djeki\'{c}}}, \bibinfo {author}
  {\bibnamefont{{H.-J. Kluge}}}, \bibinfo {author}
  {\bibfnamefont{S.}~\bibnamefont{Stahl}}, \bibinfo {author}
  {\bibfnamefont{T.}~\bibnamefont{Valenzuela}}, \bibinfo {author}
  {\bibfnamefont{J.}~\bibnamefont{Verd\'{u}}}, \bibinfo {author}
  {\bibfnamefont{M.}~\bibnamefont{Vogel}},\ and\ \bibinfo {author}
  {\bibfnamefont{G.}~\bibnamefont{Werth}},\ }%
  \bibfield{journal}{%
  \bibinfo {journal} {Nucl. Inst. Meth. Phys. Res. B}\ }%
  \textbf{\bibinfo {volume} {214}},\ \bibinfo {pages} {207} (\bibinfo {year}
  {2004})%
  \bibAnnoteFile{NoStop}{QuintAntiprotonAspirations}%
\bibitem{MainzSummary2006}%
  \BibitemOpen
  \bibfield{author}{%
  \bibinfo {author} {\bibfnamefont{G.}~\bibnamefont{Werth}}, \bibinfo {author}
  {\bibfnamefont{J.}~\bibnamefont{Alonso}}, \bibinfo {author}
  {\bibfnamefont{T.}~\bibnamefont{Beier}}, \bibinfo {author}
  {\bibfnamefont{K.}~\bibnamefont{Blaum}}, \bibinfo {author}
  {\bibfnamefont{S.}~\bibnamefont{Djekic}}, \bibinfo {author}
  {\bibfnamefont{H.}~\bibnamefont{H\"affner}}, \bibinfo {author}
  {\bibfnamefont{N.}~\bibnamefont{Hermanspahn}}, \bibinfo {author}
  {\bibfnamefont{W.}~\bibnamefont{Quint}}, \bibinfo {author}
  {\bibfnamefont{S.}~\bibnamefont{Stahl}}, \bibinfo {author}
  {\bibfnamefont{J.}~\bibnamefont{Verd\'u}}, \bibinfo {author}
  {\bibfnamefont{T.}~\bibnamefont{Valenzuela}},\ and\ \bibinfo {author}
  {\bibfnamefont{M.}~\bibnamefont{Vogel}},\ }%
  \bibfield{journal}{%
  \bibinfo {journal} {Int. J. Mass Spectrom.}\ }%
  \textbf{\bibinfo {volume} {251}},\ \bibinfo {pages} {152} (\bibinfo {year}
  {2006})%
  \bibAnnoteFile{NoStop}{MainzSummary2006}%
\bibitem{DehmeltMagneticBottle}%
  \BibitemOpen
  \bibfield{author}{%
  \bibinfo {author} {\bibfnamefont{R.}~\bibnamefont{{Van Dyck, Jr.}}}, \bibinfo
  {author} {\bibfnamefont{P.}~\bibnamefont{Ekstrom}},\ and\ \bibinfo {author}
  {\bibfnamefont{H.}~\bibnamefont{Dehmelt}},\ }%
  \bibfield{journal}{%
  \bibinfo {journal} {Nature}\ }%
  \textbf{\bibinfo {volume} {262}},\ \bibinfo {pages} {776} (\bibinfo {year}
  {1976})%
  \bibAnnoteFile{NoStop}{DehmeltMagneticBottle}%
\bibitem{Review}%
  \BibitemOpen
  \bibfield{author}{%
  \bibinfo {author} {\bibfnamefont{L.~S.}\ \bibnamefont{Brown}}\ and\ \bibinfo
  {author} {\bibfnamefont{G.}~\bibnamefont{Gabrielse}},\ }%
  \bibfield{journal}{%
  \bibinfo {journal} {Rev. Mod. Phys.}\ }%
  \textbf{\bibinfo {volume} {58}},\ \bibinfo {pages} {233} (\bibinfo {year}
  {1986})%
  \bibAnnoteFile{NoStop}{Review}%
\bibitem{Entanglement2008}%
  \BibitemOpen
  \bibfield{author}{%
  \bibinfo {author} {\bibfnamefont{R.}~\bibnamefont{Blatt}}\ and\ \bibinfo
  {author} {\bibfnamefont{D.~J.}\ \bibnamefont{Wineland}},\ }%
  \bibfield{journal}{%
  \bibinfo {journal} {Nature}\ }%
  \textbf{\bibinfo {volume} {453}},\ \bibinfo {pages} {1008} (\bibinfo {year}
  {2008})%
  \bibAnnoteFile{NoStop}{Entanglement2008}%
\bibitem{CompareAlAndHgIonClocks2008}%
  \BibitemOpen
  \bibfield{author}{%
  \bibinfo {author} {\bibfnamefont{T.}~\bibnamefont{Rosenband}}, \bibinfo
  {author} {\bibfnamefont{D.~B.}\ \bibnamefont{Hume}}, \bibinfo {author}
  {\bibfnamefont{P.~O.}\ \bibnamefont{Schmidt}}, \bibinfo {author}
  {\bibfnamefont{C.~W.}\ \bibnamefont{Chou}}, \bibinfo {author}
  {\bibfnamefont{A.}~\bibnamefont{Brusch}}, \bibinfo {author}
  {\bibfnamefont{L.}~\bibnamefont{Lorini}}, \bibinfo {author}
  {\bibfnamefont{W.~H.}\ \bibnamefont{Oskay}}, \bibinfo {author}
  {\bibfnamefont{R.~E.}\ \bibnamefont{Drullinger}}, \bibinfo {author}
  {\bibfnamefont{T.~M.}\ \bibnamefont{Fortier}}, \bibinfo {author}
  {\bibfnamefont{J.~E.}\ \bibnamefont{Stalnaker}}, \bibinfo {author}
  {\bibfnamefont{S.~A.}\ \bibnamefont{Diddams}}, \bibinfo {author}
  {\bibfnamefont{W.~C.}\ \bibnamefont{Swann}}, \bibinfo {author}
  {\bibfnamefont{N.~R.}\ \bibnamefont{Newbury}}, \bibinfo {author}
  {\bibfnamefont{W.~M.}\ \bibnamefont{Itano}}, \bibinfo {author}
  {\bibfnamefont{D.~J.}\ \bibnamefont{Wineland}},\ and\ \bibinfo {author}
  {\bibfnamefont{J.~C.}\ \bibnamefont{Bergquist}},\ }%
  \bibfield{journal}{%
  \bibinfo {journal} {Science}\ }%
  \textbf{\bibinfo {volume} {319}},\ \bibinfo {pages} {1808} (\bibinfo {year}
  {2008})%
  \bibAnnoteFile{NoStop}{CompareAlAndHgIonClocks2008}%
\bibitem{SidebandCoolingOfNeutralAtoms1998}%
  \BibitemOpen
  \bibfield{author}{%
  \bibinfo {author} {\bibfnamefont{H.}~\bibnamefont{Perrin}}, \bibinfo {author}
  {\bibfnamefont{A.}~\bibnamefont{Kuhn}}, \bibinfo {author}
  {\bibfnamefont{I.}~\bibnamefont{Bouchoule}},\ and\ \bibinfo {author}
  {\bibfnamefont{C.}~\bibnamefont{Salomon}},\ }%
  \bibfield{journal}{%
  \bibinfo {journal} {Europhys. Lett.}\ }%
  \textbf{\bibinfo {volume} {42}},\ \bibinfo {pages} {395} (\bibinfo {year}
  {1998})%
  \bibAnnoteFile{NoStop}{SidebandCoolingOfNeutralAtoms1998}%
\bibitem{SidebandCoolingMechanicalResonator}%
  \BibitemOpen
  \bibfield{author}{%
  \bibinfo {author} {\bibfnamefont{T.}~\bibnamefont{Rocheleau}}, \bibinfo
  {author} {\bibfnamefont{T.}~\bibnamefont{Ndukum}}, \bibinfo {author}
  {\bibfnamefont{C.}~\bibnamefont{Macklin}}, \bibinfo {author}
  {\bibfnamefont{J.~B.}\ \bibnamefont{Hertzberg}}, \bibinfo {author}
  {\bibfnamefont{A.~A.}\ \bibnamefont{Clerk}},\ and\ \bibinfo {author}
  {\bibfnamefont{K.~C.}\ \bibnamefont{Schwab}},\ }%
  \bibfield{journal}{%
  \bibinfo {journal} {Nature}\ }%
  \textbf{\bibinfo {volume} {463}},\ \bibinfo {pages} {72} (\bibinfo {year}
  {2010})%
  \bibAnnoteFile{NoStop}{SidebandCoolingMechanicalResonator}%
\bibitem{WinelandFrequencyDivision}%
  \BibitemOpen
  \bibfield{author}{%
  \bibinfo {author} {\bibfnamefont{D.~J.}\ \bibnamefont{Wineland}},\ }%
  \bibfield{journal}{%
  \bibinfo {journal} {J. Appl. Phys.}\ }%
  \textbf{\bibinfo {volume} {50}},\ \bibinfo {pages} {2528} (\bibinfo {year}
  {1979})%
  \bibAnnoteFile{NoStop}{WinelandFrequencyDivision}%
\bibitem{VanDyckMagnetronCoolingLimit}%
  \BibitemOpen
  \bibfield{author}{%
  \bibinfo {author} {\bibfnamefont{R.~S.}\ \bibnamefont{{Van Dyck, Jr.}}},
  \bibinfo {author} {\bibfnamefont{P.~B.}\ \bibnamefont{Schwinberg}},\ and\
  \bibinfo {author} {\bibfnamefont{H.~G.}\ \bibnamefont{Dehmelt}},\ }%
  \enquote{\bibinfo {title} {New frontiers in high energy physics},}\ \
  (\bibinfo {publisher} {{(Plenum, New York)}},\ \bibinfo {year} {1978})\ p.\
  \bibinfo {pages} {159}%
  \bibAnnoteFile{NoStop}{VanDyckMagnetronCoolingLimit}%
\bibitem{QuintDoublePenningTrap}%
  \BibitemOpen
  \bibfield{author}{%
  \bibinfo {author} {\bibfnamefont{H.}~\bibnamefont{H\"affner}}, \bibinfo
  {author} {\bibfnamefont{T.}~\bibnamefont{Beier}}, \bibinfo {author}
  {\bibfnamefont{S.}~\bibnamefont{Djeki\'c}}, \bibinfo {author}
  {\bibfnamefont{N.}~\bibnamefont{Hermanspahn}}, \bibinfo {author}
  {\bibfnamefont{H.-J.}\ \bibnamefont{Kluge}}, \bibinfo {author}
  {\bibfnamefont{W.}~\bibnamefont{Quint}}, \bibinfo {author}
  {\bibfnamefont{S.}~\bibnamefont{Stahl}}, \bibinfo {author}
  {\bibfnamefont{J.}~\bibnamefont{Verd\'u}}, \bibinfo {author}
  {\bibfnamefont{T.}~\bibnamefont{Valenzuela}},\ and\ \bibinfo {author}
  {\bibfnamefont{G.}~\bibnamefont{Werth}},\ }%
  \bibfield{journal}{%
  \bibinfo {journal} {Eur. Phys. J. D}\ }%
  \textbf{\bibinfo {volume} {22}},\ \bibinfo {pages} {163} (\bibinfo {year}
  {2003})%
  \bibAnnoteFile{NoStop}{QuintDoublePenningTrap}%
\bibitem{OneIonTemperature}%
  \BibitemOpen
  \bibfield{author}{%
  \bibinfo {author} {\bibfnamefont{S.}~\bibnamefont{Djekic}}, \bibinfo {author}
  {\bibfnamefont{J.}~\bibnamefont{Alonso}}, \bibinfo {author}
  {\bibnamefont{{H.-J. Kluge}}}, \bibinfo {author}
  {\bibfnamefont{W.}~\bibnamefont{Quint}}, \bibinfo {author}
  {\bibfnamefont{S.}~\bibnamefont{Stahl}}, \bibinfo {author}
  {\bibfnamefont{T.}~\bibnamefont{Valenzuela}}, \bibinfo {author}
  {\bibfnamefont{J.}~\bibnamefont{Verd\'u}}, \bibinfo {author}
  {\bibfnamefont{M.}~\bibnamefont{Vogel}},\ and\ \bibinfo {author}
  {\bibfnamefont{G.}~\bibnamefont{Werth}},\ }%
  \bibfield{journal}{%
  \bibinfo {journal} {Eur. Phys. J. D.}\ }%
  \textbf{\bibinfo {volume} {31}},\ \bibinfo {pages} {451} (\bibinfo {year}
  {2004})%
  \bibAnnoteFile{NoStop}{OneIonTemperature}%
\bibitem{PbarMass}%
  \BibitemOpen
  \bibfield{author}{%
  \bibinfo {author} {\bibfnamefont{G.}~\bibnamefont{Gabrielse}}, \bibinfo
  {author} {\bibfnamefont{X.}~\bibnamefont{Fei}}, \bibinfo {author}
  {\bibfnamefont{L.~A.}\ \bibnamefont{Orozco}}, \bibinfo {author}
  {\bibfnamefont{R.~L.}\ \bibnamefont{Tjoelker}}, \bibinfo {author}
  {\bibfnamefont{J.}~\bibnamefont{Haas}}, \bibinfo {author}
  {\bibfnamefont{H.}~\bibnamefont{Kalinowsky}}, \bibinfo {author}
  {\bibfnamefont{T.~A.}\ \bibnamefont{Trainor}},\ and\ \bibinfo {author}
  {\bibfnamefont{W.}~\bibnamefont{Kells}},\ }%
  \bibfield{journal}{%
  \bibinfo {journal} {Phys. Rev. Lett.}\ }%
  \textbf{\bibinfo {volume} {65}},\ \bibinfo {pages} {1317} (\bibinfo {year}
  {1990})%
  \bibAnnoteFile{NoStop}{PbarMass}%
\bibitem{OpenTrap}%
  \BibitemOpen
  \bibfield{author}{%
  \bibinfo {author} {\bibfnamefont{G.}~\bibnamefont{Gabrielse}}, \bibinfo
  {author} {\bibfnamefont{L.}~\bibnamefont{Haarsma}},\ and\ \bibinfo {author}
  {\bibfnamefont{S.~L.}\ \bibnamefont{Rolston}},\ }%
  \bibfield{journal}{%
  \bibinfo {journal} {Intl. J. of Mass Spec. and Ion Proc.}\ }%
  \textbf{\bibinfo {volume} {88}},\ \bibinfo {pages} {319} (\bibinfo {year}
  {1989}),\ \bibinfo {note} {ibid. {\bf 93}, 121 (1989)}%
  \bibAnnoteFile{NoStop}{OpenTrap}%
\bibitem{PbarMassPRL1ppb}%
  \BibitemOpen
  \bibfield{author}{%
  \bibinfo {author} {\bibfnamefont{G.}~\bibnamefont{Gabrielse}}, \bibinfo
  {author} {\bibfnamefont{D.}~\bibnamefont{Phillips}}, \bibinfo {author}
  {\bibfnamefont{W.}~\bibnamefont{Quint}}, \bibinfo {author}
  {\bibfnamefont{H.}~\bibnamefont{Kalinowsky}}, \bibinfo {author}
  {\bibfnamefont{G.}~\bibnamefont{Rouleau}},\ and\ \bibinfo {author}
  {\bibfnamefont{W.}~\bibnamefont{Jhe}},\ }%
  \bibfield{journal}{%
  \bibinfo {journal} {Phys. Rev. Lett.}\ }%
  \textbf{\bibinfo {volume} {74}},\ \bibinfo {pages} {3544} (\bibinfo {year}
  {1995})%
  \bibAnnoteFile{NoStop}{PbarMassPRL1ppb}%
\bibitem{ElectronCalorimeter}%
  \BibitemOpen
  \bibfield{author}{%
  \bibinfo {author} {\bibfnamefont{D.~J.}\ \bibnamefont{Wineland}}\ and\
  \bibinfo {author} {\bibfnamefont{H.~G.}\ \bibnamefont{Dehmelt}},\ }%
  \bibfield{journal}{%
  \bibinfo {journal} {J. Appl. Phys.}\ }%
  \textbf{\bibinfo {volume} {46}},\ \bibinfo {pages} {919} (\bibinfo {year}
  {1975})%
  \bibAnnoteFile{NoStop}{ElectronCalorimeter}%
\end{thebibliography}
%\end{document}

%Merlin.mbs v4.21 2009-07-09.
%

\end{document}